\documentclass[twocolumn]{aastex6}
\usepackage{graphicx,txfonts}

\usepackage{threeparttablex,longtable,booktabs}

\shortauthors{Pineda \& Hallinan}
\shorttitle{Radio Limit for TRAPPIST-1}

\begin{document}
\title{A Deep Radio Limit for the TRAPPIST-1 System}
\author{J. Sebastian Pineda\altaffilmark{1} and Gregg Hallinan\altaffilmark{2}}



\affil{\altaffilmark{1} University of Colorado Boulder, Laboratory for Atmospheric and Space Physics, 3665 Discovery Drive, Boulder CO, 80303, USA; \href{mailto:sebastian.pineda@lasp.colorado.edu}{sebastian.pineda@lasp.colorado.edu}  }

\affil{\altaffilmark{2} California Institute of Technology, Department of Astronomy, 1200 E. California Ave, Pasadena CA, 91125, USA}

\begin{abstract}
The first nearby very-low mass star planet-host discovered, TRAPPIST-1, presents not only a unique opportunity for studying a system of multiple terrestrial planets, but a means to probe magnetospheric interactions between a star at the end of the main sequence and its close-in satellites. This encompasses both the possibility of persistent coronal solar-like activity, despite cool atmospheric temperatures, and the presence of large-scale magnetospheric currents, similar to what is seen in the Jovian system. Significantly, the current systems include a crucial role for close-in planetary satellites analogous to the role played by the Galilean satellites around Jupiter. We present the first radio observations of the seven-planet TRAPPIST-1 system using the \textit{Karl G. Jansky Very Large Array}, looking for both highly circularly polarized radio emission and/or persistent quiescent emissions. We measure a broadband upper flux density limit of $<$8.1 $\mu$Jy across 4-8 GHz, and place these observations both in the context of expectations for stellar radio emission, and the possible electrodynamic engines driving strong radio emissions in very-low mass stars and brown dwarfs, with implications for future radio surveys of TRAPPIST-1 like planet-hosts. We conclude that magnetic activity of TRAPPIST-1 is predominantly coronal and does not behave like the strong radio emitters at the stellar/sub-stellar boundary. We further discuss the potential importance of magnetic field topology and rotation rates, demonstrating that a TRAPPIST-1 like planetary system around a rapidly rotating very-low mass star can generate emission consistent with the observed radio luminosities of very-low mass stars and brown dwarfs.
\end{abstract}

\keywords{brown dwarfs}

\section{Introduction}\label{sec:intro}

Recently, \cite{Gillon2017} announced the detection of a seven-planet system, including three Earth-sized planets in the liquid water habitable zone, orbiting a nearby very-low mass star, TRAPPIST-1. The system is particularly exciting because it provides appealing targets for ongoing and future transmission spectroscopy observations, and a way to conduct comparative exoplanetology in terrestrial bodies around a single host star \citep[e.g.,][]{Barstow2016, Morley2017, deWit2018}. This system also probes planet formation around the lowest mass stars \citep[e.g.,][]{Alibert2017}, and will be a benchmark for characterizing exoplanets with properties similar to Earth \citep{Gillon2017}. 

Interestingly, TRAPPIST-1 resides in the regime of ultracool dwarfs (UCDs; spectral type $\gtrsim$ M7) at the end of the main sequence. These objects exhibit a transition with cooler effective temperature from coronal stellar activity, as is seen on more massive stars, to magnetic emissions driven by large-scale magnetospheric currents like those that power the multi-wavelength aurorae of Jupiter as well as some brown dwarfs \citep{Pineda2017}. While many low-mass stars, late-type M dwarfs, exhibit the X-ray emission, stellar winds, UV transition region spectral lines, chromospheric H$\alpha$, and radio emission properties consistent with the coronal solar-like paradigm \citep[see within][]{Pineda2017}, there is a distinct sub-population of very-low mass stars whose magnetic activity indicators show very different properties, specifically, extensive evidence for rotationally pulsed electron cyclotron maser (ECM) emission, consistent with the presence of magnetospheric auroral currents \citep[see within][]{Pineda2017}. In the late L dwarf and T dwarf regime, these auroral processes dominate the magnetic activity \citep{Pineda2016, Pineda2017}. This population is most clearly evident through observations revealing much stronger (several orders of magnitude) radio emission from these objects than would be expected based on the empirical G\"{u}del-Benz relation connecting coronal X-ray and radio emissions in low-mass stars \citep{Gudel1993, Williams2014}, as well as the presence of periodic highly circularly polarized emission. Although it is unclear what the underlying conditions are that drive the presence of large-scale magnetospheric currents in these objects, it is likely related to rapid rotation rates, and strong large-scale magnetic fields \citep{Pineda2017, Turnpenney2017}. The presence of auroral radio emission might then be related to the bistability of magnetic dynamos in fully convective low-mass stars as seen through Zeeman Doppler Imaging (ZDI), which show either strong dipolar large-scale fields or weaker non-axisymmetric multi-polar fields \citep{Morin2010}, with the strongest average field strengths observed among the fastest rotating objects with predominantly dipolar field topologies \citep{Shulyak2017}. If auroral radio processes require these large-scale dipolar fields, then the aurorae could be indicative of which branch of the dynamo regime a given object may inhabit. Moreover, the prevalence of a given fully-convective dynamo branch would then influence the statistics of auroral radio detections. Understanding this connection could provide a new means to probe fully-convective dynamos in low-mass stars.

Fast rotation and high field strengths may be necessary conditions to drive significant auroral magnetospheric currents, however, the nature of the underlying electrodynamic engine powering the current system in these UCDs remains an open question. Possible hypotheses include magnetospheric-ionospheric coupling currents driven by the large-scale motions of an equatorial plasma disk \citep{Schrijver2009, Nichols2012, Turnpenney2017}, reconnection between the large-scale magnetosphere and the interstellar medium \citep{Turnpenney2017}, or the electrodynamic interaction between the UCD's magnetic field and any nearby orbiting satellites \citep{Hallinan2015,Pineda2017}. Models of these processes suggested that they can generate sufficient energy to power the observed emissions \citep{Zarka2007,Schrijver2009,Saur2013,Nichols2012,Turnpenney2017}, but it remains unclear which of these mechanisms, or a mixture of them, are predominately responsible for generating the auroral ECM emission of some UCDs. Similarly, the origins of the quiescent radio emission in this same population of radio UCDs is uncertain. This emission is likely synchrotron or gyrosynchrotron radiation and may be associated with high energy electrons trapped in closed magnetospheric loops, akin to the Jovian radiation belts \citep{Hallinan2006, Pineda2017}. However, the continual mass-loading of the magnetosphere with plasma, a requirement for both the existence of the plasma radiation belts and several of the proposed electrodynamics engines is itself an open question, and might be related to volcanic planetary activity, similar to the Jovian system, or possibly atmospheric sputtering \citep{Hallinan2015}.

As the only known nearby UCD planet-host, TRAPPIST-1 provides the first opportunity to test whether the presence of a close-in exoplanetary system plays a significant role in generating these magnetic emission processes around very-low mass stars and brown dwarfs, possibly through a direct interaction with the stellar host or by providing the magnetospheric plasma source. By observing TRAPPIST-1 at radio wavelengths, we can look for the possibility of electron cyclotron maser emission or quiescent radio emission, and assess whether its activity properties (e.g., radio, X-ray, and H$\alpha$) are consistent with the coronal solar-like paradigm or whether it may belong to the sub-population of very-low mass stars exhibiting auroral phenomena. Understanding how this stellar system, with potentially significant interactions with its planetary satellites, fits into the transition in magnetic activity indicators in the UCD regime thus motivates the current radio study of this very-low mass star. This letter is organized as follows. In Section~\ref{sec:targ}, we review the properties of TRAPPIST-1. In Section~\ref{sec:data}, we discuss our data set from the \textit{NSF's Karl G. Jansky Very Large Array} \citep[\textit{VLA};][]{Perley2011}. In Section~\ref{sec:model}, we compare our observations to estimates of the stellar radio flux densities from TRAPPIST-1. In Section~\ref{sec:discuss}, we discuss the role of satellites for possible auroral emission mechanisms in UCDs. Lastly, in Section~\ref{sec:summary}, we summarize our findings with implications for future searches of radio emission at the end of the main sequence.

\begin{ThreePartTable}

\begin{TableNotes}
	\item[a]{Rotation period is taken from \textit{K2} photometric variability.}
	\item[b]{Denotes the surface averaged magnetic field strength from Zeeman Broadening measurements; $f$ is the filling factor between 0 and 1.}
	\item[]{R\textsc{eferences.} -- (1) \citealt{Burgasser2015}; (2) \citealt{vanGrootel2018}; (3) \citealt{Vida2017}; (4) \citealt{Reiners2018}; (5) \citealt{Reiners2010}; (6) \citealt{Wheatley2017}; (7) \citealt{Bourrier2017}; (8) \citealt{Burgasser2017}; (9) This Paper }
\end{TableNotes}

\begin{longtable}{l c c } 
	\caption{Properties of the UCD TRAPPIST-1} \label{tab} \\
	\toprule
	Property & Value & Reference  \\
	\midrule
	\endfirsthead
	
	\toprule
	Property & Value & Reference  \\
	\midrule
	\endhead

	\bottomrule
	\endfoot
	
	\bottomrule
	\insertTableNotes
	\endlastfoot

	Spectral Type \dotfill & M8 & (1) \\ [3pt]   
	Distance \dotfill & $12.14 \pm 0.12$ pc & (2) \\ [3pt]  
	Mass \dotfill & $0.089\pm0.006$ $M_{\odot}$& (2) \\[3pt] 
	Radius \dotfill & $0.121\pm0.003$ $R_{\odot}$ & (2) \\[3pt] 
	$L_{\mathrm{bol}}$ \dotfill  & $5.22 \pm 0.19 \times 10^{-4}$ $L_{\odot}$& (2) \\[3pt] 
	$T_{eff}$ \dotfill & $2516 \pm 41 $ K& (2) \\ [3pt] 
	Rotation Period\tablenotemark{a} \dotfill & $3.295\pm 0.003$ days & (3) \\ [3pt] 
	$v \sin i$ \dotfill & $< 2$ km s$^{-1}$ & (4) \\ [3pt] 
	$B f$ \tablenotemark{b} \dotfill & 600 $\pm^{200}_{400}$ G & (5) \\ [3pt] 
	$L_{\mathrm{X}}$ \dotfill & 3.8-7.9 $\times$  $10^{26}$ erg s$^{-1}$ & (6) \\[3pt] 
	$\log_{10}(L_{\mathrm{Ly}\alpha} / L_{\mathrm{bol}})$ \dotfill & $\sim -4.1$& (7) \\[3pt] 
	$\log_{10}(L_{\mathrm{H}\alpha} / L_{\mathrm{bol}})$ \dotfill & $\sim -4.7$& (8) \\[3pt] 
	$L_{\nu}$ (4-8 GHz ) \dotfill & $<$1.43 $\times$ 10$^{12}$ erg s$^{-1}$ Hz$^{-1}$ & (9)\\[3pt]
\end{longtable}

\end{ThreePartTable}

\begin{figure*}[htbp]
	\centering
	\includegraphics[width=0.9\textwidth]{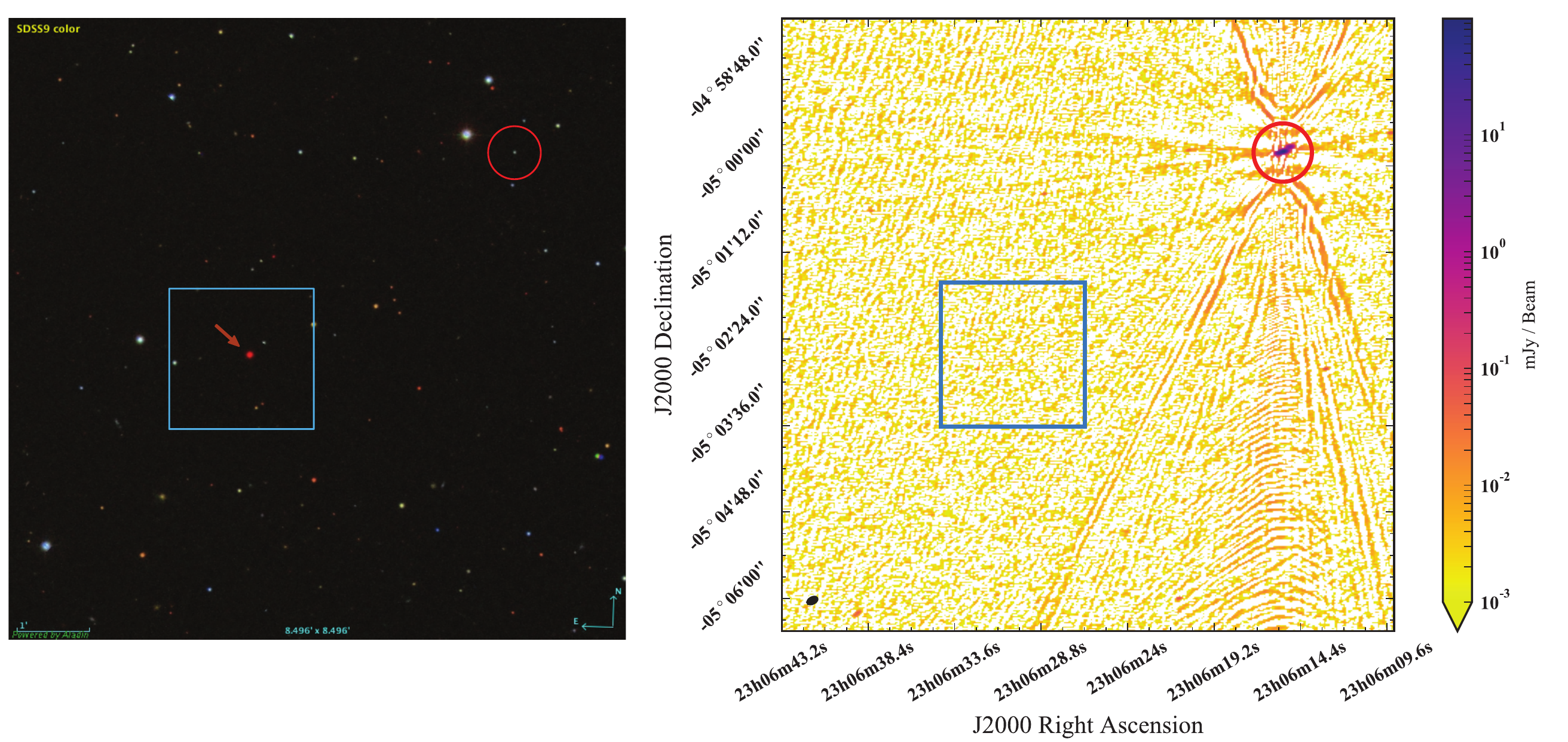} 
	\caption{\textit{Left} - Multi-color 8.5'x8.5' optical image of the field near TRAPPIST-1 using SDSS DR9 photometry \citep{Ahn2012}, made using \textit{Aladin} \citep{Bonnarel2000}. The square is centered on the position of TRAPPIST-1 at the epoch of our radio observations, with the arrow showing TRAPPIST-1 at the image epoch, and the circle indicating the location of the nearby quasar (4C -05.95) with bright radio emission. \textit{Right} - The same 8.5'x8.5' celestial field shown at left as seen by the \textit{VLA} at 6 GHz across the full 4 GHz bandpass in Stokes I. The square and circle show the expected location of TRAPPIST-1 and the quasar in the radio field of view, respectively. To reach the thermal noise sensitivity in these observations the removal of the quasar source needed to be treated very carefully, see Section~\ref{sec:data}. Note that the colorbar is on a logarithmic scale to show both the thermal noise floor (2.7~$\mu$Jy per beam) and the bright quasar. The size and shape of the synthesized beam is shown in the lower left.}
	\label{fig:images}
\end{figure*}

\section{TRAPPIST-1}\label{sec:targ}

TRAPPIST-1, also know as 2MASS J23062928--0502285, is an M8 dwarf at 12.1 pc which hosts seven terrestrial planets, detected in transit from photometric monitoring \citep{Gillon2016, Gillon2017, Luger2017}. The star has observed variable H$\alpha$ and Ly$\alpha$ emission with typical levels of $\log_{10}(L_{\mathrm{H}\alpha}/L_{\mathrm{bol}}) \sim -4.7$ \citep{Burgasser2017} and $\log_{10}(L_{\mathrm{Ly}\alpha}/L_{\mathrm{bol}}) \sim -4.1$ \citep{Bourrier2017}, respectively, and displays photometric variability at optical wavelengths from \textit{K2} monitoring with a periodicity of 3.295~d \citep{Luger2017, Vida2017}. This updated rotation period differs from the initially published period in \cite{Gillon2016} of 1.4~d, but is consistent with the updated projected rotational velocity of $v \sin i < 2 $ km~s$^{-1}$ \citep{Reiners2018}\footnote[3]{There was some tension in these measurements initially with $v \sin i = 6 \pm 2 $ km~s$^{-1}$ \citep{Reiners2010}.}; \cite{Roettenbacher2017} discussed the discrepancy in the period measurements, attributing the different results to changing stellar surface features (also see \citealt{Morris2018}, who consider the possibility of bright surface spots generating the $\sim$3.3 d periodogram signal). \textit{XMM Newton} observations have also detected an X-ray luminosity of (3.8-7.9) $\times 10^{26}$ erg s$^{-1}$ in the band 0.1-2.4 keV, which is consistent with observations of other late M dwarfs, although amongst the strongest such emitters \citep{Williams2014, Wheatley2017}. The stellar physical properties and emission characteristics of TRAPPIST-1 are summarized in Table~\ref{tab}, using the values based on new parallax measurements from \cite{vanGrootel2018}.

\begin{figure*}[htbp]
	\centering
	\includegraphics[width=0.9\textwidth]{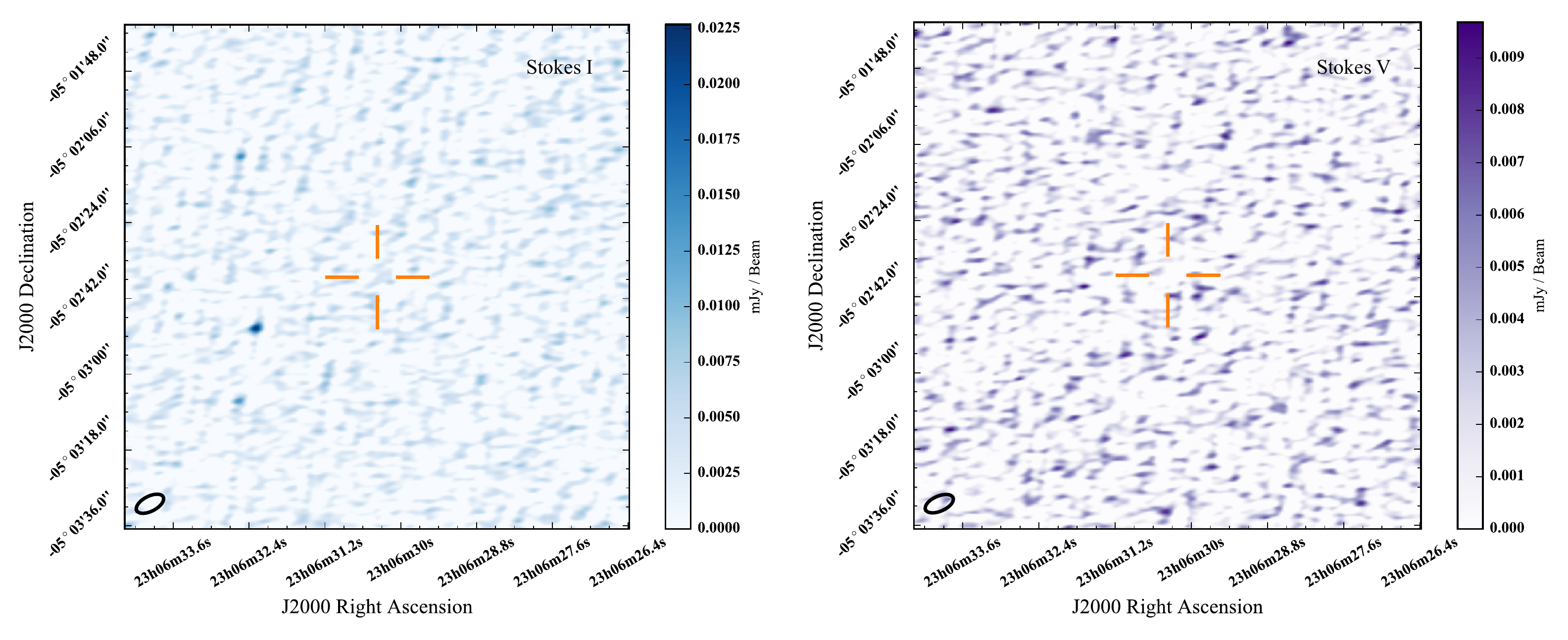} 
	\caption{\textit{Left} - A zoom in view of the Stokes I radio image around the expected location of TRAPPIST-1, corresponding to the 2' wide squares in Figure~\ref{fig:images}. We do not detect a source at this position and report a 3$\sigma$ upper limit of $<$8.1~$\mu$Jy.
	\textit{Right} - The zoom in view of the Stokes V radio image around the expected location of TRAPPIST-1, corresponding to the same 2' wide squares shown in Figure~\ref{fig:images}. We do not detect any evidence for circularly polarized emission from TRAPPIST-1.}
	\label{fig:stokes}
\end{figure*}

\section{OBSERVATIONS} \label{sec:data}

In order to categorically rule out periodic bright radio pulsations from TRAPPIST-1, we would need to monitor the star for a duration exceeding its rotation period, as well as the orbital period of the inner planets. However, in the case of all known periodically pulsing UCDs, the pulsed emission is accompanied by a quiescent radio component, favoring a short initial search for quiescent emission as a proxy for the ECM induced emissions \citep{Pineda2017}. We thus conducted an initial pilot observation of TRAPPIST-1 with the \textit{VLA} (project 16A-466, PI Pineda) on 2016 May 12 from UT 12:57:37 to 14:57:16 in the CnB configuration, to look for circularly polarized pulses and measure a potential quiescent radio component. We chose to observe at C-band (4--8~GHz) due to the success of previous volume-limited radio surveys in detecting UCD radio emission between 4 and 10 GHz \citep{Antonova2013}, as well as evidence that the quiescent radio luminosity of this population broadly peaks in this band \citep{Ravi2011, Williams2015c}. Observations were carried out using the 4~GHz bandwidth enabled by the 3-bit sampler mode of the the WIDAR correlator.

After initial setup scans, the flux calibrator 3C48 was observed, followed by observations of the target interleaved with short observations of a standard gain calibrator, the quasar QSO B2320-035, every 10 minutes. The total resulting time on the target was $\sim$90 minutes. Initial data editing, RFI excision and calibration was carried out using Common Astronomy Software Applications (CASA) VLA calibration pipeline \citep{McMullin2007}. Imaging the resulting calibrated data revealed a 100~mJy source $\sim$4 arcminutes from the position of TRAPPIST-1, limiting the rms noise to $\sim$30~$\mu$Jy per beam in the region of TRAPPIST-1. This bright quasar source is evident in Figure~\ref{fig:images}. Three iterations of phase only self-calibration followed by two subsequent iterations of phase and amplitude self calibration resulted in rms noise in the region of TRAPPIST-1 of 2.7~$\mu$Jy per beam in Stokes I and 2.6~$\mu$Jy per beam in Stokes V, consistent with expectation for thermal noise.

 We measure a flux density level of 3.8 $\mu$Jy in the synthesized beam at the proper motion corrected position for TRAPPIST-1, consistent with the noise level. We show radio images in Stokes I and Stokes V at the location of TRAPPIST-1 in Figure~\ref{fig:stokes}, with no source clearly evident above the noise level. We thus report a 3$\sigma$ flux density upper limit for TRAPPIST-1 from 4-8 GHz of $<$8.1 $\mu$Jy for our unresolved source, which corresponds to a specific luminosity of $<$1.43 $\times$ $10^{12}$ erg~s$^{-1}$ Hz$^{-1}$. We performed an additional search for short duration pulses of emission by using the CASA task \textit{fixvis} to shift the phase center of the data to the proper motion corrected position for TRAPPIST-1 and subtracting a model of all other sources in the field using the CASA task \textit{uvsub}. We then plotted the real part of the complex visibilities, averaged across all baselines, frequencies and both polarizations, representing the light curve at the location of TRAPPIST-1. No evidence for emission from TRAPPIST-1 was present in the light curve. Finally, we imaged each of the nine individual 10~minute scans on the target in Stokes I and V to search for emission on these timescales, with no evidence of any source above the 3$\sigma$ detection threshold.

\section{Stellar Emission Mechanisms} \label{sec:model}

Our observations provide the first radio flux density limits from TRAPPIST-1, allowing us to compare this result to possible expectations from different models of the magnetic activity from this star.
 
\subsection{Coronal Emission}\label{sec:gudel}
 
The strong X-ray emissions from TRAPPIST-1 suggest that it possesses a hot coronal atmosphere. Based on the known X-ray emission, the G\"{u}del-Benz relation, using the linear fit to the literature data from \cite{Berger2010}, would predict a radio luminosity of only 2.2-6.0 $\times$ 10$^{10}$ erg~s$^{-1}$~Hz$^{-1}$, below our measured limit of $<$10$^{12.2}$ erg~s$^{-1}$~Hz$^{-1}$, and consistent with the solar-like paradigm of magnetic activity. Our radio limit is also an order of magnitude lower then the typical quiescent radio luminosity of the known radio emitting M7-M9.5 dwarfs, which depart significantly from the G\"{u}del-Benz relation \citep{Williams2014,Pineda2017}. TRAPPIST-1 is likely very similar to other very low-mass stars that show coronal behavior, like VB~8 or VB~10 \citep{Hawley2003}, which have comparable limits on their radio flux densities and exhibit X-ray emission \citep{Berger2008}. Future radio observations, however, would require an improvement of two orders of magnitude in thermal noise to confirm whether the G\"{u}del-Benz relation continues to describe the X-ray and radio luminosities of these very-low mass stars.

Nevertheless, these findings are in agreement with the conclusions of \cite{Stelzer2012} suggesting that amongst ultracool dwarfs, the X-ray bright and radio loud objects are distinct populations \citep[see also ][]{Williams2014}. Building on results from \cite{McLean2012}, \cite{Williams2014} proposed a magnetic-field topology difference between the two populations, based on the possibility of multiple dynamo modes in this stellar regime \citep{Morin2010}. \cite{Williams2014} additionally based their conclusions on the continuity from mid-M dwarf studies showing a correlation between average surface magnetic field measurements and X-ray luminosity. Interestingly, \cite{Shulyak2017} showed that for low-mass stars with measured topologies from ZDI, either largely multi-polar or dipolar, both types can display strong X-ray emission ($\log_{10} L_{X} / L_{\mathrm{bol}} \sim -3$), with average surface fields up to 4 kG in the multi-polar case, and even larger in the dipolar case. Extending this into the UCD regime, these results would suggest that the X-ray luminosity alone cannot be used to separate the populations with different large-scale topologies. \cite{Williams2014} attributed the X-ray bright objects to the population with axisymmetric dipolar large-scale fields and the radio load objects to those with predominantly multi-polar fields. We posit that it is more likely to be the reverse, as the mechanisms theorized to produce the radio emission require strong dipolar large-scale field topologies \citep{Pineda2017,Turnpenney2017}, and in light of the \cite{Shulyak2017} results, that the X-ray emission alone cannot be used to distinguish the likely topology, the radio measurements are necessary to classify the X-ray bright and radio loud populations. Moreover, if the presence of the radio emission depends on additional factors, not just the field topology or stellar properties (see Section~\ref{sec:discuss}), the populations of X-ray and radio emitting objects may not be totally mutually exclusive. There is some evidence of this already in the small handful of radio UCDs with detected quiescent X-ray luminosities \citep{Williams2014}. Accordingly, these data could be explained if strong ($\gtrsim$4 kG surface averaged) large-scale dipolar fields are a requisite to power the radio emission, but significant X-ray emissions are possible with either topology, depending on other properties, such as $T_{eff}$. In this scenario for UCDs, X-ray emission measurements would not provide any information of the large-scale field topology, but the presence of radio emission would. An X-ray non-detection may suggest a weaker average surface field without constraining the topology, and a radio non-detection would similarly leave the question of topology open. ZDI measurements of the population of radio loud UCDs, confirming or refuting whether their large-scale fields are predominately axisymmetric and dipolar, would significantly help determine the relationship between these various measures of magnetic activity, and assess how the topology may differ between the populations posed by \cite{Stelzer2012}.

\subsection{Stellar Wind-Planet Interaction}\label{sec:wind}

Our radio flux density limits also provide new constraints on the physical parameters determining the power of its potential radio emission. Using the framework of \cite{Saur2013}, \cite{Turnpenney2018} demonstrated that the planets around TRAPPIST-1 could induce significant radio emission by serving as an obstacle to a magnetized stellar wind flow from the host star. In this sub-Alfv\'{e}nic interaction, the Poynting flux dissipated across the planetary obstacle in the wind flow propagates toward the stellar surface in Alfv\'{e}n waves that power the electron cyclotron maser instability \citep{Turnpenney2018}. We note that this scenario is not seen in the solar system, as the magnetized solar system planets interacting with the solar wind are in a super-Alfv\'{e}nic regime \citep{Zarka2007}\footnote[4]{The super- and sub-Alfv\'{e}nic regimes are determined by whether the wind speed exceeds the Alfv\'{e}n speed of the local magnetic field.}. Their results indicate that TRAPPIST-1 could emit steady-state radio emission $\sim$10 $\mu$Jy with possible bursts two orders of magnitude greater due to stochastic processes in the wind flow, magnetic field strength, and/or field orientation \citep{Turnpenney2018}. However, assumptions in this calculation, for example, concerning the stellar wind outflow rate or the planetary magnetic field strengths are very uncertain, and thus might be overestimating the true levels of radio emission generated. Nevertheless, with our measured flux density limit of $<$8.1 $\mu$Jy at 6 GHz, we do not observe any radio emission consistent with these estimates, and thus our observations begin to constrain the parameter space available within the TRAPPIST-1 system to generate radio emission through this interaction between the stellar wind and the close-in planets. However, a parameter space search is beyond the scope of this work, as the combined modeling assumptions concerning the form of the magnetized stellar wind, planetary magnetic fields, and radio emission properties make constraints on the individual parameters, like the wind outflow rate or planetary dipole moment, inconsequential without additional data constraining these processes. Within the 2 hr duration of our observations, we also do not see evidence for any possible bursts, providing a constraint on the duty cycle of the possible stochastic events that might generate bursts of radio emission.

However, an important consideration is that the ECM emission modeled by \cite{Turnpenney2018} is emitted at frequencies tied to the magnetic field strength in the vicinity of the source, presumably near the stellar surface, which has an average surface field strength of only 600~G \citep{Reiners2010}, corresponding to 1.68 GHz; the lower bound of our passband at 4 GHz corresponds to $\sim$1.43 kG fields. This nominal field strength, $Bf$, averages together both magnetic and non-magnetic regions across the stellar surface, and thus the highly magnetized regions, such as above star spots, likely hold much stronger fields, $\sim$600$/f~\mathrm{G}$. However, the filling factor is unknown in the Zeeman Broadening measurements of mid-late M dwarfs as $B$ and $f$ are not separable \citep{Reiners2007}, making the maximum surface field strengths uncertain. Although the average field strength corresponds to frequencies lower than our observing band, ECM emission may still be generated up to the maximum field strengths on the stellar surface, possibly encompassed by our 4-8 GHz observations. If there were ECM source regions tied to field lines with these higher field strengths, we may have expected to see this kind of emission based on the estimates from \cite{Turnpenney2018}. The lack of a detection could be explained by a paucity of magnetic field regions exceeding strengths $\sim$1.5 kG, our observing frequencies being potentially too high to probe the weaker magnetic fields of TRAPPIST-1. A ZDI map of the this star's magnetic field topology would help determine whether this was indeed the case. Alternatively, even if these regions were prevalent, there is no guarantee that these also corresponded to ECM source regions. A deep search at lower frequencies than our observations is warranted to rule out or potentially detect ECM generated from this star-planet interaction. 

Additionally, due to the beaming of the ECM emission into less than 4$\pi$ sr \citep{Zarka2004, Treumann2006}, our observations may not have been optimally oriented in space or time to intercept the emission. Although the beaming geometry of hollow and wide ECM source cones near the rotational axis favors observability when viewed near an inclination of 90$^{\circ}$ \citep{Pineda2017}, which is likely the case if the planetary system angular momentum and magnetic axes are aligned with the stellar rotation axis, the rotational and orbital phases need to be well sampled to rule out ECM emission entirely --- the 2 hr extent of our observations is much smaller than the several day rotation and orbital periods of TRAPPIST-1 and its planetary system.

\subsection{Auroral Emission}\label{sec:Ha}

Although we did not see any highly polarized pulsations, the limited rotational phase coverage ($\sim$3\%) in our observations cannot rule them out, nevertheless our measured upper limit on the flux density of quiescent radio emission does provide information on the likelihood of possible GHz ECM emission from TRAPPIST-1, given the statistical association of quiescent and pulsed emission amongst UCDs \citep{Pineda2017}. Although the underlying cause is uncertain, \cite{Pineda2017} demonstrated a correlation between the observed quiescent radio luminosity at GHz frequencies, likely of synchrotron origin \citep{Ravi2011, Williams2015b}, and H$\alpha$ luminosity among known periodically pulsing radio UCDs. If TRAPPIST-1 behaved like those auroral objects, we would expect a radio luminosity of $\sim$10$^{13.8}$ erg s$^{-1}$ Hz$^{-1}$, but instead our measured limit of $<$10$^{12.2}$ erg s$^{-1}$ Hz$^{-1}$ is over a magnitude smaller, well below that relationship even accounting for the scatter of $\sim$0.24 dex, at fixed H$\alpha$ luminosity. Even if TRAPPIST-1 exhibited radio emission at levels below our detection threshold, the star would inhabit a region of $L_{\mathrm{H}\alpha}-L_{\nu,\; \mathrm{rad}}$ space with other very low-mass stars that have their H$\alpha$ luminosities dominated by chromospheric emission, instead of being associated with the presence of auroral currents \citep{Pineda2017}. Our results thus confirm that H$\alpha$ emission is not a viable proxy for quiescent radio emission in the TRAPPIST-1 system, it is not auroral and instead is likely dominated by chromospheric emission, and that TRAPPIST-1 is not likely to generate radio emission in the same manner as the known auroral radio UCDs.

\section{Role of Planets in Producing UCD Radio Emission}\label{sec:discuss}

Our aim in conducting these observations was to provide a deep limit for quiescent radio emission from the TRAPPIST-1 system in the context of both stellar radio emission and auroral processes in the UCD regime. We further aimed to test whether the presence of a planetary system is a crucial ingredient to the production of strong\footnote[5]{By `strong' we are referring to radio emissions exceeding the G\"{u}del-Benz relation predictions by several orders of magnitude, which for UCDs could be $L_{\nu} \sim10^{13.5}$ erg s$^{-1}$ Hz$^{-1}$, although there is a broad range, see Section~\ref{sec:gudel} and \cite{Williams2014}.} radio emission from UCDs, either through a direct star-planet interaction that produces ECM emission or by the detection of quiescent radio emission associated with the presence of magnetospheric currents possibly due to equatorial radiation belts. While our null detection in this instance points toward the coronal paradigm for TRAPPIST-1 and leaves the question of possible star-planet interactions open, if there is a crucial role for close-in planetary companions, our results point to the necessity of multiple conditions that must be met to produce strong radio emitters (G\"{u}del-Benz deviants, see Section~\ref{sec:gudel}) in the population of very-low mass stars and brown dwarfs. Such close-in satellites may or may not be necessary, but their presence is certainly not sufficient to drive the electrodynamic engines of radio UCDs.

\subsection{Magnetospheric Mass-Loading}

All of the proposed hypotheses for the electrodynamic engine driving UCD radio emission (see Secton~\ref{sec:intro}), except reconnection with the ISM, require a significant source of plasma internal to the system. Similarly, the radiation belts that might explain the quiescent emission of these systems requires the magnetosphere to be loaded with plasma. Using the Jovian system as an example, a volcanic planet, like Io, can be this source. However, our radio observations show no evidence for these magnetospheric plasma structures around TRAPPIST-1. This could be a consequence of either of two distinct possibilities, first, the planetary satellites are not providing sufficient plasma to the magnetosphere, or the magnetosphere does not sustain large-scale loops with which to contain the plasma. Even though the equilibrium tidal heating of the TRAPPIST-1 planets from N-body simulations suggest internal heat fluxes comparable or potentially greater than that of Io \citep{Luger2017}, the volcanism on these planets may not be contributing to the mass-loading of the magnetosphere. One reason may be that the larger masses of the TRAPPIST-1 planets \citep{Wang2017, Grimm2018} relative to the Galilean moons prevent significant amounts of volcanic material from escaping the planetary atmospheres. Even if a strong stellar wind, as suggested by \cite{Garraffo2017}, can erode the planetary atmospheres, the material would likely be entrained with the wind along the open field lines, instead of populating a steady state plasma torus like that observed in the Jovian magnetosphere \citep[e.g.,][]{Bagenal1997}. Any material that does populate the magnetospheric environment must be trapped in closed large-scale magnetic-loops in order to create an equatorial radiation belt of synchrotron emission. Without a significant large-scale component to the stellar magnetic field, the ability of such stars to generate very strong quiescent radio emission similar to that observed from the known radio UCDs may be limited, irrespective of any mass-loading.

\subsection{Jupiter-Io Analogues in UCD Systems}\label{sec:Jup}

Our radio observations of TRAPPIST-1 also provide a test of the possibility of magnetic  flux tube interactions analogous to the Jupiter-Io system powering radio emission from UCDs. Our non-detection suggests these processes, if powering the known radio UCDs, are not taking place in the TRAPPIST-1 system, or are too weak to detect. One reason for this is likely associated with the slow TRAPPIST-1 rotation rate, relative to known radio UCDs, as well as possibly the role of large-scale magnetic field topologies. This is evident in models of the sub-Alfv\'{e}nic interaction driving the Jupiter-Io current system responsible for Io-related Jovian decametric ECM radio emission \citep{Saur2013}. The relevant equation for the total surface integrated Poynting flux\footnote[6]{For convenience and to reflect its origins, we refer to the output of Equation~\ref{eq:poynt} as the `Poynting flux', however the quantity has units of luminosity.} generated by the differential motion of a planetary body embedded in a large magnetosphere in the limit of small Alfv\'{e}n mach numbers (small velocities relative to the Alfv\'{e}n speed) from \cite{Saur2013} is 

\begin{equation}
S_{\mathrm{total}} = \frac{1}{2}\bar{\alpha}^{2} R_{o}^{2}  \varv^{2} B \sin^{2} \theta \sqrt{4\pi \rho}   \, ,
\label{eq:poynt}
\end{equation}

\noindent where $\bar{\alpha}$ is the dimensionless interaction strength\footnote[7]{$\bar{\alpha} \sim$0.5 for the Galilean satellites \citep{Saur2013} but could be near unity for the TRAPPIST-1 planets \citep{Turnpenney2018}}, $R_{o}$ is the effective radius of the planet obstacle defined by the planetary magnetosphere or ionosphere, $B$ is the magnetic field strength from the star in the vicinity of the planet in Gaussian units, $\varv$ is the relative speed of the planet through the stellar magnetosphere, $\theta$ is the angle between the magnetic field and the relative velocity vector, and $\rho$ is the mass density of the plasma environment. While the presence of a planet provides the necessary obstacle, and potentially supplies a sufficiently dense plasma environment, a strong stellar generated magnetic field and rapidly rotating magnetosphere are also necessary.

To illustrate this, we consider, for the innermost planet of the TRAPPIST-1 system, the Poynting flux generated according to Equation~\ref{eq:poynt}, if the star hosted a large-scale dipolar field consistent with its measured average surface field (see Table~\ref{tab}). We take $\bar{\alpha}\rightarrow 1$, $\theta=90^{\circ}$, and plasma densities similar to values around Io, with a number density of $\sim$2000 cm$^{-3}$ and mean molecular weight of 22 amu \citep{Saur2013}. The radius of the obstacle is taken as at minimum, the planetary radius 1.127 R$_{\oplus}$ of TRAPPIST-1b \citep{Delrez2018}. The velocity is the relative velocity between the orbital motion and the rotating magnetosphere at the location of the planet, $\varv \sim 45$ km s$^{-1}$. The field strength at the location of the planet is $\sim$0.05 G, assuming the planet lies along the magnetic equator. Plugging in these values gives a Poynting flux of $\sim2.5\times 10^{13}$ W. Assuming a conversion efficiency of 1\% from Poynting flux to radio power, a beam solid angle of 1.6 sr \citep{Zarka2004, Turnpenney2018}, and using a 4 GHz bandwidth consistent with EMC observations of UCDs \citep{Hallinan2015}, this would correspond to a radio flux density of $\sim$0.03~$\mu$Jy, well below current radio observatory capabilities. This flux density drops further if the 1.4~d period is used since that is very near the orbital period of TRAPPIST-1b. If instead we consider the same planet in a 1~d orbit around an UCD like TRAPPIST-1 but with a 2~hr rotation period, like that observed from radio UCDs \citep{Pineda2017}, and 5~kG average surface field strength with a dipolar large-scale field topology consistent with ZDI measurements \citep{Shulyak2017}, the corresponding flux density of ECM emission would be $\sim$300 $\mu$Jy ($L_{\nu,\,\mathrm{rad}} = 10^{12.6}$ erg s$^{-1}$ Hz$^{-1}$), readily detectable. Although this estimate is smaller than the strength of some of the observed highly circularly polarized radio bursts from UCDs (see Table~1 of \citealt{Pineda2017}), it is subject to many unknown quantities, including the radio emission efficiency of the ECM instability ($\sim$0.01), and the beaming solid angle ($\sim$1.6~sr), in addition to other system properties like the plasma environment. For example, while a beaming solid angle of $\sim$1.6~sr is commonly used as a basis for estimating ECM radio fluxes from UCDs \citep{Nichols2012, Turnpenney2017}, it could be as low as $\sim$0.16~sr \citep{Queinnec2001,Zarka2004}, which would increase the predicted flux by a factor of 10. Our estimates based on Equation~\ref{eq:poynt} are linear in the magnetic field strength, so a factor of 10 weaker field (500 G), with the same rapid rotation could still produce detectable emission ($\sim$30 $\mu$Jy ), however, if the rotation is slightly slower or the plasma environment is less dense, the prospects for currently detectable emission become marginal. Additionally, if the field is mostly multi-polar, the field strength at the planet location would drop off more quickly with distance, further limiting the strength of these potential emissions. Nevertheless, even if TRAPPIST-1 generated ECM emission like the Jupiter-Io system, assuming an optimistic large-scale field topology, it would have been too weak to detect with our current observations. Several kG surface magnetic field strengths in dipolar topologies, and fast rotation rates ($\sim$2 hr) are necessary to generate currently detectable GHz frequency ECM radio emission through a star-planet flux tube interaction, like that of Jupiter and Io. Conversely, if the TRAPPIST-1 planetary system was hosted by one of the rapidly rotating UCDs with strong ($\gtrsim$4 kG surface averaged) large-scale dipolar magnetic fields, then the system would generate ECM emission consistent with the emission levels observed in many of the known radio UCDs. Given the estimates of planet occurrence rates in short period orbits around very-low mass stars \citep[$\sim$30\%;][]{He2017}, and the rapid rotation rates of most UCDs with likely strong dipolar fields, the overall radio UCD detection rate may be plausibly determined by the presence of these three conditions.

\section{Conclusions}\label{sec:summary}

Our observations of the TRAPPIST-1 system centered at 6~GHz yielded no detectable radio emission to a limit of $<$8.1~$\mu$Jy. In Section~\ref{sec:gudel}, we demonstrated that this limit was consistent with the G\"{u}del-Benz relation, applicable to coronally active low-mass stars, and motivated future ZDI observations of radio loud UCDs to discern the role that magnetic field topology plays in dictating the activity indicators in this stellar population. In Section~\ref{sec:wind}, we compared our radio measurements to the possible strength of electron cyclotron maser emission driven by an impinging stellar wind, as calculated by \cite{Turnpenney2018}, concluding that we did not see any such ECM emission from the TRAPPIST-1 system, possibly due to low rotational/orbital phase coverage, with further observations, including at lower frequencies, required to rule out the possibility of wind driven ECM emission. In Section~\ref{sec:Ha} we also compared this limit to possible quiescent emission levels based on correlations among the population of known radio ultracool dwarfs \citep{Pineda2017}, illustrating that TRAPPIST-1 does not likely exhibit detectable radio emission that behaves in the same manner as the emission from the known periodically pulsing radio UCDs.

Although we can not use these new data on TRAPPIST-1 to discern the role, if any, that planetary systems have in generating bright radio emissions in the population of very low-mass stars and brown dwarfs, it is evident that the presence of a planetary system by itself does not guarantee strong radio emission at GHz frequencies. If the observed ECM radio emissions of UCDs are generated through a magnetic interaction analogous to the Jupiter-Io system (see Section~\ref{sec:Jup}), additional criterion beyond the presence of a close-in planet must also be met, namely rapid rotation and strong (several kG surface averaged) magnetic field strengths --- the TRAPPIST-1 system does not satisfy these later two conditions. Additionally, even if planets are present around UCDs, their capacity to provide a source of magnetospheric plasma is still an open question depending on the tidal heating and volcanism of those planets. Similarly, the retention of such a magnetospheric plasma likely depends on the magnetic field topology and the presence of large-scale magnetic loops. These loops may then serve as the site for the persistent quiescent synchrotron emission. 

In light of the ZDI observations of fully convective low-mass stars showing two regimes of dynamos and their respective large-scale field topologies, either predominantly dipolar or multi-polar, with the strongest fields associated with the dipolar topology and fastest rotators \citep{Shulyak2017}, the connection between observed ECM and quiescent radio emission could be explained as a coherent consequence of this strong large-scale dipolar field when a sufficient plasma source is available. The kG magnetic field strengths help power the UCD auroral electrodynamic engine \citep{Turnpenney2017} and the closed large-scale field houses the magnetospheric plasma that generates the quiescent emission. These assorted criteria may then collectively contribute to the low detection statistics for UCD radio emission \citep[e.g.,][]{Lynch2016}. Future surveys looking for UCD radio emission at GHz frequencies are more likely to succeed targeting the fastest rotating objects, while the slower rotators or objects with significant X-ray emission may be better targeted at lower frequencies, hundreds of MHz. TRAPPIST-1, as a slow rotator, with strong coronal X-ray and weak radio emissions, and likely possessing a multi-polar large-scale field, reflects the population of very-low mass stars with coronal solar-like activity instead of the sub-population exhibiting auroral magnetic processes. We suggest that the X-ray emission of a given object alone is not a sufficient indication of the likely field topology, but in conjunction with radio emission measurements can provide an indication of that topology to the extent that it is responsible for the dichotomy of observed X-ray and radio properties amongst UCDs. This is potentially powerful as ZDI measurements of these objects are difficult with current instrumentation due to the faint intrinsic luminosities and rapid rotation of UCDs. Nevertheless, these assessments remain circumstantial; new measurements of the magnetic field topology of UCDs are necessary to definitively establish this connection between the magnetic emissions and the field topology.

It remains to be seen whether the strong ECM and quiescent radio emissions of the few radio UCDs is related to the presence of planets, or are predominantly driven without them \citep{Turnpenney2017}. While the magnetic field strength and rotation rate of TRAPPIST-1 are consistent with the non-detection of radio emission within a Jupiter-Io flux tube paradigm, it is notable that the same planetary configuration orbiting a rapidly rotating dwarf with a large-scale dipolar field can account for the observed radio luminosities of radio emitting UCDs. The prevalence of multi-planet systems in tight orbits orbiting UCDs, as well as the fraction of rapid rotators with strong magnetic fields, may prove to be consistent with the detection rate of radio pulsed emission from these systems. However, such evidence would be circumstantial with direct confirmation requiring detection of radio pulses from a TRAPPIST-1 like system. More deep radio searches in soon to be discovered planetary systems around very-low mass stars and brown dwarfs, as well as potential planet detections among the growing population of radio detected UCDs will help elucidate the answer.

\section*{Acknowledgments}

J.S.P would like to thank Kevin France, Zach Berta-Thompson and Jackie Villadsen for helpful comments in the preparation of this manuscript. GH acknowledges the generous support of NSF Career award AST-1654815. The authors would also like to thank the anonymous referee for the thoughtful consideration of this manuscript and for providing comments that strengthened this article. 

The National Radio Astronomy Observatory is a facility of the National Science Foundation operated under cooperative agreement by Associated Universities, Inc. 

Funding for SDSS-III has been provided by the Alfred P. Sloan Foundation, the Participating Institutions, the National Science Foundation, and the U.S. Department of Energy Office of Science. The SDSS-III web site is http://www.sdss3.org/.

SDSS-III is managed by the Astrophysical Research Consortium for the Participating Institutions of the SDSS-III Collaboration including the University of Arizona, the Brazilian Participation Group, Brookhaven National Laboratory, Carnegie Mellon University, University of Florida, the French Participation Group, the German Participation Group, Harvard University, the Instituto de Astrofisica de Canarias, the Michigan State/Notre Dame/JINA Participation Group, Johns Hopkins University, Lawrence Berkeley National Laboratory, Max Planck Institute for Astrophysics, Max Planck Institute for Extraterrestrial Physics, New Mexico State University, New York University, Ohio State University, Pennsylvania State University, University of Portsmouth, Princeton University, the Spanish Participation Group, University of Tokyo, University of Utah, Vanderbilt University, University of Virginia, University of Washington, and Yale University. 

\software{Aladin \citep{Bonnarel2000}, CASA \citep{McMullin2007}}

\bibliographystyle{aasjournal}
\bibliography{trappist}

\end{document}